# Giulio Racah and Theoretical Physics in Jerusalem

Nissan Zeldes[a)]

The Racah Institute of Physics, The Hebrew University of Jerusalem, Jerusalem 91904, Israel

**Contents:**

**Abstract:** Giulio Racah's name is known world wide in physics due to the mathematical methods he developed that are based on tensor operators and continuous groups. These methods revolutionized spectroscopy, and they are currently essential research tools in atomic, nuclear and elementary particle physics. He himself applied them to modernizing theoretical atomic spectroscopy. Racah laid the foundations of theoretical physics in Israel. He educated several generations of Israeli physicists, and put Israel on the world map of physics.

## I. Introduction. Racah in Italy

Forty-two years have passed since the untimely death of Giulio Racah (1909-1965). The present article considers Racah's contributions to general physical theory and his establishment of theoretical physics as a discipline in Israel. The reader is assumed to have the required knowledge of non-relativistic quantum mechanics, elementary angular momentum and group representation theory. This Introduction briefly summarizes Racah's scientific career in Italy before emigrating to Palestine. His later contributions in Jerusalem are addressed in the following sections. The beginnings of physics in Jerusalem before Racah are related in ref.[1].

Racah became a graduate student in physics shortly after the discovery of Quantum Mechanics, at a time when modern physics began flourishing in Italy. Under the impetus of Corbino research groups were formed around Enrico Fermi in Rome and Enrico Persico in Florence, with close contact between the two. Most prominent in the Florence department were Bruno Rossi, Giuseppe Occhialini, Gilberto Bernardini and Giulio Racah, who remained life-long close friends[2, 3]. Racah's doctoral thesis (1930) on a treatment of light interference phenomena by Fermi's quantum theory of radiation (no. 2 in Talmi's list[4]) was written under Persico's supervision. He thereafter stayed one year (1930-31) as Fermi's assistant in Rome and another (1931-32) with Pauli in Zürich. Upon his return to Italy he first taught theoretical Physics at the university of Florence (1932-1937) and thereafter in Pisa (1936-37). After coming out second in a nation-wide competition for a chair in theoretical physics in Palermo (1937) (See Emilio Segrè's autobiography[5]) he was appointed professor (Professore Straordinario) in Pisa, the youngest professor there.

Most of Racah's work in Italy was in the new domains opened by quantum theory. Best known are his calculations of the cross section for bremsstrahlung (nos. 12 and 13 in Talmi's list) and for electron-positron pair production (nos. 19 and 20) in the collisions of fast charged particles. The calculations were very involved, but Racah succeeded in obtaining analytical formulas where others obtained only approximate ones. The results are important for cosmic ray research. Another well known work of his (no. 25) clarifies the transformation properties under reflections of relativistic spin ½ particles obeying the Dirac equation, with applications to Majorana's theory of the neutrino. These early works already demonstrated Racah's outstanding mathematical and computational abilities and the thoroughness of his approach, which became characteristic features of his work. He also calculated the expected hyperfine structure and isotope shifts of atomic spectral lines (nos. 5, 6 and 8), discussed the symmetry properties of tensors using group theory (9-11, 23, 24) and kept abreast of developments in nuclear physics (7, 16, 17, 26). Pauli and Fermi appreciated him deeply. When the anti-Semitic racial laws were passed in Italy (1938) Racah was dismissed from the University of Pisa and applied for a professorial position at The Hebrew University of Jerusalem. The excellent letters of reference written on his behalf by Fermi and Pauli were decisive for his appointment[1].

Racah indicated[6] (see also [7]) that his choice of Jerusalem was due to a consultation with Dr. Augusto Levi, head of the Italian Zionist Organization, who stressed the magnitude of the challenge of creating theoretical physics in Jerusalem where there was none, as compared to a less demanding effort in an already existing department somewhere else. The support and encouragement obtained from the head of the World Zionist Organization, Dr. Chaim Weizmann[1], gave Racah the final required stimulus to decide to go to Jerusalem.

## II. World War II and the Israel War of Independence. Racah establishes his home in Jerusalem

Racah was appointed full professor of theoretical physics at the Hebrew University of Jerusalem in November 1939, and became the youngest professor there. He immigrated together with his mother, and was determined to become integrated fast in the new land and the University. He studied Hebrew intensively, and after half a year mastered it to the extent that he could start participating in the meetings of the Senate of the University[1]. Since the teaching at the University is done in Hebrew he regularly attended the General Physics course given by Dr. Shmuel Sambursky in order to learn the required Hebrew terminology[8]. A year after his arrival (1940-41) he already taught in Hebrew the basic theoretical courses:

Analytical Mechanics, Electromagnetic Theory and Quantum Theory[9]. Additionally he held a weekly theoretical physics seminar. This teaching load was higher than what he was used to in Italy[1], and is also considerably heavier than presently accepted standards.

During 1945-46 Racah was the head of the Teaching Committee of the Faculty of Science[10], and during 1946-48 he was Dean of the Faculty[11].

Let me briefly mention some aspects of his personal life: In September 1940 Racah married Zmira Many, a daughter of the Supreme Court judge Malchiel Many. They had a daughter (1941) and two sons (1944, 1949).

In 1942 he joined the main pre-state clandestine military organization Haganah. He took a course for squad commanders and became an instructor in the use of small arms and field exercises[12]. At the outbreak of the War of Independence in November 1947 he was deputy commander of the Haganah forces on Mount Scopus[12, 13] which was then (and remained so until the end of the Six Days War in June 1967) a Jewish enclave surrounded by Arab territory. From March 1948 till the end of the War he was in command of a scientific unit in Jerusalem preparing munitions for the army[12, 14, 15]. He also designed conical hollow charges which were used against the wall of the Old City of Jerusalem when the attempt was made to reconquer it in July 1948[12, 15].

All this while Racah did not neglect his scientific research, which he considered his vocation. In the year 1940, when he adopted a new homeland, learned a new language and started a family, he also started working on his major scientific work, The Theory of Complex Spectra.

Racah thus cemented his roots in Israel. He declined an offer made to him by the Italian government before the end of World War II[10] to resume his university position in Pisa and told Bernardini[3] that he would have doubts in himself if he returned. But almost every Summer, on his way to scientific conferences after the strains of the academic year in Jerusalem, he would rest, meet friends and regain strength in Florence.

Racah's deep commitment to his new homeland and to his students is obvious in his emotive Introductory words to the book Analytical Mechanics[16], written from notes taken during his lectures by physics second year undergraduates Israel Grinfeld and Yoram Treves. Both of them were killed early in 1948 in the War of Independence, fighting in battles for the control of the highway to Jerusalem, then under siege:

"Introductory words

In the summer of 1947 two excellent students, Israel Grinfeld and Yoram Treves, approached me offering to publish their notes of my lectures on Analytical Mechanics. Having known both of them, and realizing how much my students are hampered in their studies by lack of suitable textbooks on the subject, I welcomed the suggestion.

I received the first parts of the manuscript in the fall of 1947, whereas the last parts were handed during hasty visits between convoys in the following winter. As well as being first in studies, they were also among the first who were deprived of their young lives defending the road to eternal Jerusalem. Israel Grinfeld fell in Sha'ar Hagai on February 29, 1948, and Yoram Treves fell near Hulda on March 31.

The friends of Israel Grinfeld and Yoram Treves ז"ל considered it their holy duty editing the manuscript and preparing it for publication, thus perpetuating the memory of the two youths, who dedicated their lives to their people and homeland. At the end of the War, while still in service, the friends started working on it under my guidance.

In addition to my lecture notes the manuscript comprised various additions and supplementary material, partly original and partly from the classical literature. We felt uneasy having to make some changes without the authors being able to defend their work and views. We attempted to reduce the changes to a minimum, in order to retain the characteristic freshness of the young authors emanating from every page.

I am unable to mention here all those who participated in scrutinizing and editing the manuscript. …I deeply thank them all.

I am sure the book will achieve its dual goal: being used as a basic text for students of Physics and Mathematics, and reminding the young students of those who sacrificed for the homeland both their lives and their bright future.

Joel Racah
Professor of Theoretical Physics
at The Hebrew University

Jerusalem, July 1950"

## III.  Theory of Complex Spectra

Racah's great original contributions to Physics are given in four classical papers "Theory of Complex Spectra I-IV"[17-20], published in the Physical Review during the years 1942-49. (See also [1].) When Racah worked on them during World War II he was isolated from the main stream of physics for five years. These works were motivated by problems arising in the study of theoretical atomic spectroscopy as described in the classical book by Condon and Shortley "The Theory of Atomic Spectra" (TAS)[21]. (According to the late Dr. Yehiel Lehrer-Ilamed of the Soreq Nuclear Research Center, Yavne, who was Racah's assistant, "Condon and Shortley" was a basic text for the theoretical seminar held by Racah in Jerusalem.)

### A.  *The central field approximation. Changing over from Slater determinants to coupled schemes*

In quantum theory as presented in TAS one calculates atomic energies and stationary states using quantum perturbation theory. In zeroth order one starts from Slater's central field approximation (1929) [TAS (Ch. VI)]. In this model the atomic electrons are assumed to move independently of one another, filling shells due to the spherical mean field around the nucleus, with single electron wave functions (wfs) $\psi_{nlm_lm_s}(\underline{r},\sigma) = \dfrac{R_{nl}(r)}{r} Y_{lm_l}(\theta,\varphi)\chi_{m_s}(\sigma)$,

where $\underline{r}$ and $\sigma$ are the respective space and spin coordinates. The single electron wf has a radial quantum number n, orbital angular momentum quantum number $l$, and orbital and spin projective quantum numbers $m_l$ and $m_s$, denoting their corresponding z-components. Traditionally the letters, s, p, d, f … denote $l$ values of 0, 1, 2, 3… .

Due to the spherical symmetry of the potential an electron's energy depends on the shell it is in ($nl$ values) but not on its $m_l$ and $m_s$ values within the shell. The zeroth order energy of the whole atom, which is the sum of electronic energies, depends on the group of $nl$ values (configuration) and is degenerate in each configuration. The Pauli Exclusion Principle requires having atomic wfs which are antisymmetric with respect to exchange of space and spin coordinates of any two electrons. The simplest such wfs are Slater determinants constructed from the single electron wfs. As a rule, for Slater determinants the total atomic angular momenta $\mathbf{L}$ (orbital angular momentum), $\mathbf{S}$ (spin) and $\mathbf{J}$ (total angular momentum, $\mathbf{J} = \mathbf{L} + \mathbf{S}$) are not constants of the motion, and sums of determinants are required to have these angular momenta well defined.

In the higher order approximations of perturbation theory one adds residual interactions neglected thus far. There is an electric component (stemming from electrostatic interaction between electrons) which is the main part in light atoms, and a magnetic component (spin-orbit interaction) which becomes important in heavy atoms. In the first order approximation one diagonalizes the residual interaction perturbation matrix in each configuration separately, obtaining the eigenvalues as first order corrections to the degenerate zero order energies, and the eigenvectors as first order stationary states. For light atoms LS-coupling prevails and the stationary states have well defined L, S and J values, and a multiplet structure as given by the atomic vector model [TAS (Chs. VII and VIII)] which describes the experimental spectra rather well. The notation for an LS term is $^{2S+1}L$ and for an individual level it is $^{2S+1}L_J$. For heavy atoms there is jj-coupling and the stationary states have well defined single-electron angular momenta $\mathbf{j} = \mathbf{l} + \mathbf{s}$ and total angular momentum $\mathbf{J} = \sum \mathbf{j}_i$ . In transition atoms, with several simultaneously filling shells, other coupling schemes occur as well[22].

Slater calculated the perturbation matrix in a basis of Slater determinants, and obtained its eigenvalues without diagonalization by using the invariance of the trace of a matrix under a transformation of basis ("the diagonal sum rule"). Thus he obtained the spectra of $\mathbf{l}_1\mathbf{l}_2$ configurations in LS-coupling and of $j_1j_2$ configurations in jj-coupling as given in TAS. For more than two electrons and higher $\mathbf{l}$ values the calculations become very lengthy. Besides, for configurations with more than one term (level) with the same LS(J) values the method gives only average energies. Spectroscopic studies by the method were restricted to simple spectra of 2-3 electrons or holes outside closed shells.

Racah maintained[23] (see also [24]) that despite its success with all two electron spectra Slater's method opened a widening unjustified gap between theoretical spectroscopy and experimental spectroscopists, because the latter were used to adopting the formulas of the vector model for interpretation, whereas the diagonal sum rule method discarded any physical model and was based on numerical tables which had to be calculated anew for each spectrum.

Due to the spherical symmetry of the intra atomic forces the total angular momentum $\mathbf{J}$ is well defined, and in many cases $\mathbf{L}$ and $\mathbf{S}$, and other resultant angular momenta of groups of electrons filling the same shells, are approximately constant too. Spectroscopic calculations in a scheme defined by coupled angular momenta are more transparent than for Slater determinants and lead to physically interpretable results like the vector model. In order to benefit from these advantages of atomic spherical symmetry Racah in the "Theory of Complex Spectra" calculated the residual interactions directly in appropriate coupled schemes which he had defined. Additionally he developed a new mathematical discipline, the algebra of tensor operators, which greatly simplifies calculations in coupled schemes.

The first problem attacked by Racah in I was the direct calculation of Slater's results for two-electron spectra in LS-coupling. The electrostatic two-electron interaction has a multipole expansion in terms of Legendre polynomials $P_k$ , $\dfrac{1}{r_{12}} = \sum_k \dfrac{r_<^k}{r_>^{k+1}} P_k\left(\cos\omega_{12}\right)$ , where $r_<$ is the lesser and $r_>$ is the greater of $r_1$ and $r_2$, and $\cos\omega_{12} = \dfrac{\mathbf{r}_1 \cdot \mathbf{r}_2}{r_1 r_2}$. For the central field single-electron wfs the two-electron electrostatic energies are differences $E^{direct} - E^{exchange}$ of respective direct and exchange energies, which are each a sum of corresponding radial integrals (Slater integrals) $F^k$ or $G^k$, multiplied by respective angular matrix elements $f_k$ or $g_k$ of $P_k(\cos\omega_{12})$ in the coupled scheme.

In I Racah obtained the matrix elements of $P_k$ by first calculating the matrix of $\cos\omega_{12}$ directly in the coupled scheme using the algebra of vector operators [TAS (Ch. III)], and then performing a direct very involved calculation of the polynomial $P_k$ of the matrix $\cos\omega_{12}$. Thus he obtained a closed formula for Slater's numerical results. He also calculated energies of simple three electron configurations $p^2 l$ and energies of nuclear $p^n$ configurations for Wigner and Majorana forces, and compared his results for $p^2 p'$ with experimental spectra of NI, OII and SII.

## B. *The algebra of tensor operators*

For more complicated three electron configurations like $d^3$ the calculation of the polynomials $P_k$ of the matrix $\cos\omega_{12}$ becomes very lengthy. Consequently Racah developed in II (and continued in III) the algebra of tensor operators and started using it in the calculations. He developed this algebra analogously to the algebra of vector operators in Ch. III of Condon and Shortley starting from the commutation relations with $\mathbf{J}$. (According to Judd[25] Racah considered this chapter one of the most remarkable in all of scientific literature.)

A cornerstone in the development of the Algebra is a function W of six angular momenta (6-j symbol) defined as a sum of products of four vector-coupling coefficients of two angular momenta [II], which also gives the transformation between different coupling schemes of three angular momenta [III]. It is called a Racah Coefficient, and the algebra is generally known as Racah Algebra, sometimes also Racah-Wigner Algebra, and more generally Algebra of Angular Momentum. How is this algebra related to Wigner? It turns out that Wigner developed it from his own point of view some years before Racah, but did not publish the results. They were included in an anthology of basic works on angular momentum in quantum mechanics, published in the nineteen-sixties[26]. In a short introduction attached to circulated copies Wigner gives credit to Racah for independently developing the subject and for recognizing its importance: "Fifty of the sixty-two pages of the present note are a verbatim copy of a manuscript that is older than the writer cares to admit. The reason that it has not been published before is, principally, that neither the writer, nor his colleagues who have read and discussed the manuscript with him, have recognized the importance of the concepts developed therein. The credit for developing most of these concepts independently, along with many other similar concepts, and for recognizing their importance, belongs principally to G. Racah, and also to H.A. Jahn, L. Biedenharn, W. Ufford, B.H. Flowers and many others. The reason that the writer's old manuscript is reproduced herewith is that he found it very useful in some recent calculations on tensor forces and that its point of view is sufficiently different from any published work to merit some interest. The preceding remarks on the origin of these notes should explain, on the other hand, the absence of more detailed references to the papers of the authors mentioned."

Angular momentum algebra is currently an essential research tool in atomic, nuclear and elementary particle physics, where there are many coupled angular momenta (see also [1]). It is addressed in any advanced text on quantum mechanics and in books devoted to it exclusively[27, 28, 29, 30]. Racah together with his cousin Ugo Fano wrote on it a monograph titled Irreducible Tensorial Sets[31]. The book was highly praised[32] for its systematic exposition, clarity and elegance, but it was not accepted as a standard text by the spectroscopic community, that seems to have preferred the earlier symbols and phase conventions[33].

After developing the Algebra Racah showed [II] that the $(2k + 1)$ spherical harmonics $Y_{kq}$ are components of an irreducible tensor of degree k, and $P_k(\cos\omega_{12})$ is a scalar product of two such tensors. He calculated its matrix elements according to the formulas of tensor algebra and obtained for the two-electron electrostatic energies in both LS- and jj-coupling closed formulas coinciding with the results obtained before. Then he discussed the one-to-one correspondence called particle-hole (p-h) conjugation between particle-states and hole-states in a many-electron shell, derived the transformation law of irreducible tensors under p-h conjugation, and calculated the term energies of the configurations $d^n$ (n = 2-5), $f^3$, $d^2p$ and $d^8p$. In the last two configurations he compared the theoretical formulas with experimental spectra of TiII and NiII. Finally he proved that $G^k/(2k+1)$ is a decreasing function of k.

## C.  Definition of a coupled scheme in a shell by fractional parentage. The seniority scheme in $d^n$ configurations

Calculations of atomic spectra using Racah techniques require having coupled schemes of antisymmetric states. For a compound configuration $\mathbf{l}_1^{n_1}\mathbf{l}_2^{n_2}$ of electrons in two shells such coupled schemes are obtainable from antisymmetric coupled schemes of the individual shells as antisymmetrized coupled products. The basis states of the individual partial configurations are called parents of the states of the compound configuration. The quantum numbers of the parents together with $LM_LSM_S$ define the compound coupled scheme [TAS (sects. $1^8$ and $2^8$), DT[34] (Chs. 25 and 37)]. Racah's calculations for the two-shell configurations $p^2\mathbf{l}$ [I], $d^2p$ and $d^8p$[II] were done in a scheme with well defined parents. On the other hand, for a configuration $\mathbf{l}^n$ of electrons in the same shell (equivalent electrons) antisymmetrization between partial groups as a rule destroys orthogonality and independence of states, and the concept of parents as defined above becomes meaningless.

In order to obtain an antisymmetric coupled scheme for equivalent electrons Racah generalized an idea of Bacher and Goudsmit (1934)[35] and developed in III a method for expanding an antisymmetric coupled state of n electrons in a shell as a sum of antisymmetric coupled states of the first $(n-1)$ electrons coupled to the $n^{th}$ one: $\left|\mathbf{l}^n\alpha LS\right\rangle = \sum_{\alpha_1L_1S_1}\left|\mathbf{l}^{n-1}(\alpha_1L_1S_1)\mathbf{l}LS\right\rangle\left\langle\mathbf{l}^{n-1}(\alpha_1L_1S_1)\mathbf{l}LS\middle|\right\}\mathbf{l}^n\alpha LS\right\rangle$, where $\alpha$ denotes quantum numbers needed additionally to $LM_LSM_S$ and the configuration for a complete definition of the states. The (n − 1)-electron states on the r.h.s. of the expansion are called fractional parents of the n-electron state, and the expansion coefficients $\left\langle\mathbf{l}^{n-1}(\alpha_1L_1S_1)\mathbf{l}LS\middle|\right\}\mathbf{l}^n\alpha LS\right\rangle$ are called coefficients of fractional parentage (cfps). This is a recursive construction of an antisymmetric coupled scheme in a shell, starting from the known two-electron states and determining at each stage the cfps for given L and S values as a solution of homogeneous linear equations ensuring that the ensuing state is antisymmetric with respect to permutations of the last (n − 1, n) electron pair. When there are several independent terms with given LS values (which happens for $\mathbf{l} \geq 2$) an arbitrary orthonormal scheme might be defined for them by using an index (ordinal number) $\alpha$ as an additional quantum number. (For the 77 terms of the configurations $d^n$ there is one triple term and fourteen double ones [III].) After defining a scheme by fractional parentage Racah obtained relations between cfps of p-h conjugate configurations and calculated formulas for matrix elements of symmetric one-electron and two-electron tensor operators using cfps.

There is still the arbitrariness associated with the definition of the scheme (the index $\alpha$) when there are multiple terms. In order to obtain general formulas this arbitrariness should be overcome. In quantum mechanics multiplicities are avoided by using a complete set of commuting observables. For the configurations $\mathbf{l}^n$ Racah added to $L^2$, $L_z$, $S^2$ and $S_z$ a scalar two-electron operator $Q = \sum q_{ij}$ counting the number of electron pairs in $^1S$ ($L = S = 0$) states.

The eigenvalues of $Q$ are given by a quantum number $v$ called seniority ($v$ comes from *vethek* (seniority in Hebrew), and $Q$ is called the seniority operator). An n-electron state of seniority $v$ is obtained from a $v$-electron state with the same quantum numbers which does not contain $^1S$ pairs, by adding $\frac{1}{2}(n-v)$ $^1S$ electron pairs and antisymmetrizing. The introduction of the seniority scheme solves the multiplicity problem for the configurations $d^n$ [III]. (For $s^n$ and $p^n$ there are no term multiplicities, and seniority is always a good quantum number which might not be introduced.)

After introducing seniority Racah calculated selection rules and other properties of the cfps and of matrix elements of irreducible tensors in the seniority scheme, and obtained the electrostatic interaction matrices between the configurations $d^n$, $d^{n-1}s$ and $d^{n-2}s^2$.

### D. Definition of a coupled scheme in a shell by irreducible representations of a chain of continuous groups. The $f^n$ configurations

For $f^n$ configurations the method of the last section for obtaining cfps is too clumsy. Besides, for $f^n$ the operators $L^2$, $L_z$, $S^2$, $S_z$ and $Q$ are not a complete set and cannot uniquely define a scheme by fractional parentage. (Of the 648 terms with well defined values of L, S and $v$ 167 terms have two- to five-fold multiplicities[36]).

Racah struggled with this problem for several years. Then he found in IV a solution based on the theory of continuous groups, which he avoided using in I-III because they were not a part of the standard equipment of theoretical physicists at that time [TAS (Introduction)]. (Racah told Pais[37] that he acquired the required knowledge of continuous groups by studying in Jerusalem during a whole year Herman Weyl's book The Theory of Groups and Quantum Mechanics[38].) He introduced in the $f^n$ configuration space a scheme of basis states belonging to unitary irreducible representations (irreps) of a chain of semi-simple continuous groups, each of which is a subgroup of the preceding group in the chain. Racah established the chain from groups of unitary transformations in the space spanned by the $2\mathbf{l} + 1$ single-electron wfs $\psi_{nlm_l}$. As the largest group he chose the unitary group $U(2\mathbf{l} + 1)$. It has as subgroups the orthogonal group $O(2\mathbf{l} + 1)$ which leaves invariant the symmetric S state $\left| \mathbf{l}^2 L = 0 \right\rangle$ of two electrons, and the group $O(3)$ of three dimensional rotations. For $\mathbf{l} \geq 2$ the corresponding group chain is $U(2\mathbf{l}+1) \supset O(2\mathbf{l}+1) \supset O(3)$.

The irreps of $U(2\mathbf{l} +1)$ are characterized by their permutational symmetry and the corresponding quantum number is S. The irreps of $O(2\mathbf{l} + 1)$ are characterized by the seniority quantum number $v$, and the irreps of $O(3)$ are characterized by L. Thus group theory leads one to the seniority scheme with basis states $\left| \mathbf{l}^n v L M_L S M_S \right\rangle$, which for the configurations $d^n$ overcomes the term multiplicity problem without additional quantum numbers $\alpha$. For $f^n$, though, $\alpha$'s are required. Racah found that for $\mathbf{l} = 3$ there is a unitary group $G_2$ (one of the five special groups in Cartan's classification in his Thesis (1894)), which is a subgroup of $O(7)$ and contains $O(3)$. He adopted a scheme corresponding to irreps of the group chain $U(7) \supset O(7) \supset G_2 \supset O(3)$ with basis states $\left| f^n \alpha \mathbf{W} \mathbf{U} L M_L S M_S \right\rangle$, where $\mathbf{W}$ is a three-component vector characterizing the irreps of $O(7)$ similarly to the seniority, $\mathbf{U}$ is a two-dimensional vector characterizing the irreps of $G_2$, and $\alpha$ denotes additional quantum numbers when required. The new scheme almost completely overcomes the term multiplicity problem for $f^n$: of the above 648 multiple terms there are only 29 pairs having the same values of $\mathbf{WULS}$[36]. They all have high excitation energies and low spins, and as a rule are unknown experimentally.

After defining the scheme Racah showed that in the above scheme the cfps are products of three factors depending each on only part of the scheme quantum numbers (quantum numbers of two neighbouring groups in the chain: $\mathbf{SW}$, $\mathbf{WU}$ and $\mathbf{U\alpha L}$), which simplifies their calculation, and gave numerical tables of the factors. Then he wrote the Slater expansion of the electrostatic energy $\sum f_k F^k$ as an expansion $\sum e_i E^i$ where the $E^i$ are linear combinations of the $F^k$ and the $e_i$ are corresponding combinations of the $f_k$, which are generalized irreducible tensors with respect to the groups $O(7)$, $G_2$ and $O(3)$. Using theorems of tensor algebra and cfp values he obtained numerical tables enabling one to calculate the electrostatic energy matrices of all $f^n$ terms.

This completed the definition of coupled schemes and the calculation of electrostatic energies for all $\mathbf{l}^n$ configurations ($s^n$, $p^n$, $d^n$ and $f^n$). It marked a dramatic breakthrough in the development of theoretical atomic spectroscopy from Slater's numerical method (determinantal wfs, the diagonal sum rule) to the comprehensive mathematical scheme used today (tensor operators, fractional parentage, chains of continuous groups, creation and annihilation operators in second quantization)[39]. The foundations of the modern approach were laid down by Racah in the Theory of Complex Spectra. De-Shalit and Talmi [DT (p. 8. Quoted in sect. VI below)] have considered him the founder of Modern Spectroscopy, both atomic and nuclear.

Racah summarized his methods on the use of group theory in spectroscopy in a series of seminar lectures he gave in spring 1951 at The Institute for Advanced Study in Princeton, where he was invited to stay on a sabbatical (1950-51) by Oppenheimer. The first three parts of the lectures address general notions on continuous groups, classification of the semi-simple groups, and their representations. The last two parts extend results from the Theory of Complex Spectra to the rapidly developing nuclear shell model[40]: The fourth part addresses the classification and calculation of nuclear states of mixed shells of neutrons and protons with isospin and seniority in LS-coupling. The fifth part treats the classification of nuclear interactions according to their tensorial character, and the calculation of the energy matrix for central interactions in $d^n$ configurations. During the following decade the symmetry approach in elementary particle physics was developed, and in hindsight it has become clear that the first three to four parts of Racah's lectures contain all the necessary tools needed for the formulation of this approach[41, 42]. The notes taken from the lectures were in demand ever since: they were mimeographed many times, reproduced as a CERN report (CERN 61-8, Geneva, 1961), and finally published in the series Springer Tracts in Modern Physics[43]. Racah himself hardly participated in the research on particle symmetries, and in the nineteen-fifties considered it "mathematical games people play with group theory"[44] (see also [41] and sect. V. C).

## IV.  Calculation and Interpretation of Atomic Spectra

### A.    *Extent of the calculations in Jerusalem*

Racah had a long-range aim of obtaining theoretical understanding of all measured atomic spectra[4, 45]. Out of 58 M.Sc. theses and 15 Ph.D. theses written under his supervision in Jerusalem, respective 46 and 8 theses comprised theoretical analyses of experimental atomic spectra using his methods[46 (7)]. Before 1954 the calculations were made using mechanical calculators (Odhner and later also Madas), and the studied spectra were relatively simple, with energy matrices that could be diagonalized with these calculators (matrices of orders 23 and 25 were already too large[24].). Due to collaboration with William Meggers, Chief of the Spectroscopy section of the Atomic and Radiation Physics Division of the National Bureau of Standards (NBS) in Washington, D.C., and Richard Trees of the NBS, it became possible for the first time in 1954 to diagonalize such matrices by the electronic computer SEAC at the Computation Laboratory of the NBS[24, 47]. In 1957 the electronic computer WEIZAC was installed at the Weizmann Institute of Science in Rehovot, and the department of theoretical physics in Jerusalem started using it routinely for theoretical analysis of spectra[48, 49]. (A lively description of the work at the WEIZAC is given by Shadmi[45].) After the WEIZAC was closed in 1963, the Department started using the PHILCO 2000 electronic computer of the Ministry of Defence, and in 1965 changed to the IBM 7040 computer newly acquired by The Hebrew University. The use of electronic computers made it possible to choose the spectra for study according to their interest rather than practical possibilities. During the nineteen-fifties and early nineteen-sixties the spectra studied in the Department were mainly complex spectra with filling d shells (the Iron, Palladium and Platinum groups)[50], and increasingly so the even more complex spectra with filling f shells in the rare earths.[51] There were also a few isolated studies of Actinide spectra.

### B. Configuration interaction

In first order perturbation theory the energies and stationary states are calculated for each configuration separately (sect. III.A). In higher orders there are additional contributions from other configurations, due to the non-diagonal part of the residual interaction between different configurations. This is called configuration interaction (CI). When the distances between the unperturbed energies of two configurations are large as compared to the magnitudes of the residual interaction matrix elements connecting the two, the perturbation series converges fast and low-order terms are sufficient. On the other hand, when the non-diagonal matrix elements are relatively large convergence is slow. Then, rather than using the perturbation series, neighbouring configurations are treated as an extended degenerate configuration [TAS (sect. $10^2$)], for which the energies and stationary states are obtained by diagonalizing an extended perturbation matrix in the extended degenerate space.

For the transition elements the pairs of states 4s and 3d, 5s and 4d and 6s and 5d have rather close single-electron energies. Therefore there is CI between close low configurations $d^n$, $d^{n-1}s$ and $d^{n-2}s^2$. Similarly, for the rare earths the states 6s, 5d and 4f have rather close single-electron energies and there is CI between low configurations $f^n d^k$, $f^n d^{k-1}s$ and $f^n d^{k-2}s^2$, and also between low configurations of the opposite parity $f^{n-1}d^{k+1}$, $f^{n-1}d^k s$ and $f^{n-1}d^{k-1}s^2$. (The parity of a configuration is even or odd according to whether $\sum l$ is even or odd. For electromagnetic forces there is CI only between configurations of the same parity.)

When many configurations interact configuration assignment to the levels becomes meaningless. Additionally there are several interactions with similar strengths, which obliterates coupling schemes. Under these circumstances the vector model is not applicable, and close cooperation between experimental and theoretical spectroscopists is called for. Racah and his collaborators fully responded to the challenge[24, 52, 53, 54, 55].

When the number of interacting configurations becomes high the density of spectral lines increases, and it becomes experimentally difficult to identify distinct energy levels by the traditional method of looking for repeated constant values of sums and differences of the spectral lines (Ritz Combination Principle (1908) [TAS (Introduction)]). This is the problem of accidental coincidences in the classification of spectral lines. The difficulty might be alleviated by taking sums and differences for pairs of a spectral line and the energy of an already known level, rather than for pairs of spectral lines, because in a dense spectrum the number of levels is much smaller than the number of lines. Racah maintained[23, 24] that the main obstacle to a complete understanding of the rare earths spectra is the paucity of known levels, and he started doing classifications together with his students[23, 56, 57]. At the same time this method started being used also at the NBS[58, 59] and at the Atomic Energy Research establishment in Harwell[60].

### C.    *Far away configurations. Effective interactions*

As a rule, diagonalization of the energy matrix in spaces comprising neighbouring configurations leaves some discrepancy between theory and experiment due to far configurations.

Such systematic deviations were first found by Trees in 1951[61, 62] in the configurations $d^5$s of Mn II and Fe III, and he accounted for them empirically by adding a correction term $\alpha L(L + 1)$ with an adjustable parameter $\alpha$ (the Trees correction). (A similar correction was found independently in Jerusalem in the M.Sc. thesis of Simcha Brudno on the lower configurations of Cr I[23, 49] but the results were not published.) Through the decade the Trees correction was found useful in many spectra (a partial summary is given in [24]).

Racah tried explaining the Trees correction as a second order perturbation[63], assuming that the correction for an n-electron configuration is the sum of separate electron pair corrections (linear theory), each of which can be accounted for exactly by adding intra-configuration two-electron effective interactions: $\alpha L(L + 1)$ for $p^2$ with three terms $^1S$, $^3P$ and $^1D$ and two Slater parameters $F^0$ and $F^2$, $\alpha L(L + 1) + \beta Q$ for $d^2$ with five terms and three Slater parameters, and $\alpha L(L+1)+\beta Q+\gamma C(G_2)$ for $f^2$ with seven terms and four parameters, where Q and $C(G_2)$ are the respective eigenvalues of the seniority operator and the Casimir operator of the group $G_2$.

Attempting to justify the linearity assumption Racah said[24] that although as a rule second order effects are not linear, Bacher and Goudsmit[35] showed that most CIs are linear to a good approximation. Later[64] he proved the linearity exactly for far away configurations obtained by two-electron excitations $l^n \rightarrow l^{n-2}l'l''$. ([64] treats the nuclear shell model in jj-coupling, but with minor modifications the proof holds for atoms in LS-coupling as well. (See also [65].))

The complete second order perturbation correction to the terms of $l^n$ due to CI with far away configurations was first calculated explicitly by Rajnak and Wybourne[66]. They showed that CI contributions due to two-electron excitations are given by expectation values in $l^n$ of two-particle intra-configuration effective operators, and single-electron excitations lead to both two- and three-particle intra-$l^n$ effective operators. Racah and his Ph.D. student Joseph Stein reached the same conclusion in a simpler and clearer way[67] by using "curtailed" operators which are essentially coupled products of creation and annihilation operators in second quantization, thus avoiding the lengthy summations over intermediate states of the perturbing configurations. For more complex excitations they obtained four-particle effective interactions as well.

The advantages of using intra-configuration effective interactions rather than the traditional inter-configuration perturbation expansion are that the order of the energy matrix does not change and only a few parameters are added.

The close agreement between theory and experiment achieved in Jerusalem since the nineteen-fifties has to a large extent been due to the consistent use of effective interactions[50, 51].

### D.   Semiempirical methods

In theoretical spectroscopy the calculation of atomic energies and stationary states is done by diagonalizing the energy matrix (sect. III.A) including effective interactions, after substituting appropriate numerical values for the radial and effective interaction parameters. The traditional method used by Racah and his collaborators for obtaining these values is the semiempirical statistical one: For a spectrum comprising N energy levels depending on the values of m independent (radial and effective) parameters, the numerical values of the parameters are determined by adjusting them in such a way that the overall mean error $\Delta$ of the calculated energies when compared to the data becomes as small as possible ("method of least squares")[48, 49]. (The mean error is given by $\Delta = \left[ \sum_i \Delta_i^2 / (N - m) \right]^{1/2}$, where $\Delta_i$ is the deviation of the i[th] calculated level from the corresponding experimental value.)

In order to obtain by this method reliable results the number N of data points should be large enough compared to the number m of adjustable parameters. For single atoms or ions this limits the analysis to spectra with sufficiently many levels. In the mid nineteen-fifties it was found that the same parameters in neighbouring atoms or ions satisfy simple (linear or quadratic) interpolation formulas as functions of the number of electrons in the shell and the degree of ionization[24, 68, 69]. This highly reduces the number of independent parameters as compared to the number of levels, and allows a simultaneous treatment of a group of atoms and ions with the same filling valence shells, with resulting high predictive power for unknown levels and even unknown complete spectra belonging to the same group. The method was developed and elaborated in Jerusalem[48, 49] and led to impressive progress in the analysis of transition elements and rare earths spectra. The difficulties encountered in the application of the iterative diagonalization – least squares procedure to complex spectra are detailed in [51].

The Landé g-factors of the levels are determined empirically from the level splittings in a magnetic field. They depend strongly on the multiplet composition of the states [TAS (sect. 3[16])] and can be used for checking the level assignments in each step of the iterations. In the last year of his life Racah planned quantifying the above use of g-factors by extending the diagonalization – least squares procedure to level energies and g-factors simultaneously. Then the problem arose of what are the appropriate relative weights for the two kinds of data. Racah gave a seminar talk about it in Jerusalem, and intended addressing it in his prospected talk at the Zeeman Centennial Conference in Amsterdam[70], as given in the book of abstracts:

"Contributed Papers, Section A

14.00-14.15    G. Racah,* Hebrew University, Jerusalem

**The Use of Zeeman Data in theoretical Calculations of Atomic Spectra**

A method has been developed for fitting by least squares to the theoretical formulas the measured g-factors of atomic levels together with the energy values. The advantages and limitations of the method will be discussed.

* Deceased August 1965."

It seems that no convincing solution to the weights problem was found, and the plan has not materialized[25].

### E.   Collaborations and international commissions

The "Theory of Complex Spectra" and the activities in Jerusalem made Racah "the principal figure in Theoretical Atomic Physics until his untimely death in 1965" (Judd)[25]. His methods and notations dominated the field. At the Rydberg Centennial Conference in Lund (1954) he delivered a key lecture on the state and problems of the theory of atomic spectra[24]. In 1955 he became a member of commission 14 (on Standards of Wave Lengths and Tables of Spectra) of the International Astronomical Union (IAU), and of the Joint Commission for Spectroscopy (JCS) of the IAU and the International Union of Pure and Applied Physics (IUPAP). In 1960 the JCS was transformed into the Triple Commission for Spectroscopy (TCS) of the IAU, IUPAP and the International Union of Pure and Applied Chemistry (IUPAC), of which Racah became a corresponding member[71]. In its Moscow meeting in 1958 the JCS decided to set up two clearing houses for atomic wfs, one for radial functions at MIT (Coster) and another for angular functions in Jerusalem (Racah), which should be able to give all information on work going on and prevent overlapping of activities by giving sufficient publicity to their existence[72]. (see also [48].) This activity continued in Jerusalem until 1963[73].

Thus the department of theoretical physics in Jerusalem became a world center for atomic spectroscopy and collaborated closely with other centers. We mentioned the collaboration with the NBS (sects. IV.A and IV.B). In the Summer of 1960 Racah was at the NBS on sabbatical and worked on rare earths spectra. In 1962 he obtained from the NBS a three-year research grant for calculations of atomic spectra, particularly for rare earths[46(3)], and he visited the NBS as an adviser[53]. The grant was extended for another year prior to its termination. At the same time he obtained a two-year grant for research on atomic spectra from the U.S. Air Force.[46(3), 65]

Another collaboration started with the CNRS in Paris. In the framework of a cultural agreement between France and Israel Racah gave at the College de France in February 1964 a series of lectures on the theory of atomic spectra[33], emphasizing the theory of configuration interaction and the special problems facing the analysis and interpretation of rare earths spectra. The lectures were highly appreciated[10] (Chabbal's letter to Racah of 17.4.1964). A collaborative research was agreed on, and by the end of 1964 Chabbal's student Yves Bordarier came to Jerusalem for three months, planning a longer visit in the following year 1965-66 when Racah would have completed his service as rector of the university.  After Racah's death Bordarier came to Jerusalem for several few-months research visits[74].

## V.   Nuclear Physics

### A.   Increasing use of the Racah Algebra

The atomic physicists were slow in adapting to the Racah methods[33]. The general recognition of their importance came from nuclear physics. After World War II, nuclear physics flourished in an increasing number of laboratories. Tensor algebra is the natural mathematical tool for treating angular correlations of successive nuclear radiations, as shown first by Racah[75] and independently by Lloyd[76]. Additionally it was found[77] that the neutrons and protons in the nucleus are filling shells similarly to the atomic electrons, and the methods developed by Racah for atomic spectroscopy are adaptable to the nuclear shell model, as shown by Racah himself for nuclear $p^n$ configurations in I but not pursued further until the revival of the nuclear shell model in the late nineteen-forties. In this connection Telegdi has recalled[78]: "The other thing I would like to say, which is perhaps more pertinent to the historian. I once asked Giulio Racah in 1950 or so why he was working on the spectrum of iron, a very esoteric thing, instead of working on nuclei. And he said to me, "You see, when I was young in 1937, Wigner and Hund wrote two fundamental papers in nuclear physics which are full of group theory, and I liked that and I studied those papers and learned the methods. But then I wanted to apply them to nuclear physics, and I found that nothing was known about nuclei. So I started applying those methods to atomic spectra where there was some evidence and I stuck with it." He had all this grandiose machinery but no nuclear levels were known. So this may be very characteristic."

Calculations of angular correlations and of nuclear structure require large numbers of numerical values of Racah coefficients and related functions. The first numerical tables[79-81] were calculated by using mechanical calculators, but shortly after that more extensive and error-free tables[82-84] were calculated by using electronic computers. The availability of comprehensive tables made possible the extension of the use of Racah's methods to various domains (including atomic spectroscopy!).

### B. Seniority schemes with isospin in the nuclear shell model

For charge-independent nuclear forces the total isospin T is well defined. It determines the permutational symmetry of the space-spin wf, and its component $T_z$ is half the nuclear neutron excess $\frac{1}{2}(N-Z)$. In a mixed shell with n neutrons and p protons the ground state (gs) value of T is as a rule the lowest possible one, namely $T_{gs} = |T_z| = \frac{1}{2}|n-p|$, which corresponds to the highest possible number of neutron-proton pairs in $T_{12} = 0$ states. This reflects the stronger attraction of a neutron-proton pair in T = 0 as compared to T = 1 states[85].

In the nucleus there is an attractive spin-orbit interaction which increases in heavy nuclei. In the light nuclei of the 1p valence shell the coupling scheme is near LS, and it changes to jj in heavier nuclei[40].

In his Princeton Lectures Racah generalized the seniority scheme from electronic $\mathbf{l}^n$ configurations in LS-coupling to mixed $\mathbf{l}^n$ shells in LST-coupling. In addition to the quantum numbers $vLM_LSM_S$ of the electronic seniority scheme there are the two isospin quantum numbers $TM_T$ and three partition quantum numbers PP'P" determining the permutational symmetry of the space and the spin-isospin wfs (respective irrep of U(2$\mathbf{l}$ + 1) and SU(4)). There is also a third isospin quantum number called reduced isospin and denoted t (see below).

Jahn and his collaborators in Southampton followed Racah's suggestion and used his generalized seniority scheme to obtain complete classification of states and numerical tables of cfps for nuclear $d^n$ and $p^n$ configurations, and the energy matrix of central-interactions for $d^n$ [86-88]. Racah calculated numerical tables enabling one to calculate the energy matrix of central interactions for $p^n$ [89] and used them together with his M.Sc. student Nissan Zeldes to explain the difference in gs spins of $^6$Li and $^{10}$B assuming intermediate coupling[90, 91].

In heavier nuclei with jj-coupling the gs spins of even-even nuclei are always $J_{gs} = 0$, and spherical odd-A nuclei have as a rule $J_{gs} = j$. This systematics corresponds to the highest possible number of identical-nucleon pairs in $J_{12} = 0$ states (the single-particle model of Mayer and Jensen[92]), reflecting the fact that the only strongly bound state of two identical nucleons in a j shell has $J_{12} = 0$ [85].

Analogously to atomic $\mathbf{l}^n$ configurations in LS-coupling one defines for nuclear $j^n$ configurations in JT-coupling a seniority operator $Q = \sum q_{ij}$ counting the number of J = 0 nucleon pairs[93, 94], and uses its eigenvalues for classifying the states.

The eigenvalues are given by two quantum numbers v and t, called respectively seniority and reduced isospin[95]. A state of n nucleons with given v and t is obtained from a state of v nucleons with the same quantum numbers $\alpha vJ$ and with T = t by adding $\frac{1}{2}(n-v)$ $J_{12} = 0$ nucleon pairs and antisymmetrizing. The basis states in the seniority scheme are $\left| j^n \beta \alpha v t J M_J T M_T \right\rangle$, where $\alpha$ and $\beta$ are quantum numbers required for a complete definition of the states in addition to $vtJM_JTM_T$ and the configuration. The $\beta$ in particular is related to the fact that $t_0$, the isospin of the added pairs, can vary between $t + T$ and $|t - T|$ according to the vector model. (For identical particles $T = \frac{n}{2}, t = \frac{v}{2}$ and $t_0 = \frac{n-v}{2}$. In this case T, t and $\beta$ are not required for the definition of the states and might not be introduced.)

Analogously to the atomic case in LS-coupling the seniority scheme for nuclei in JT-coupling is obtainable by introducing basis states belonging to irreps of a chain of unitary transformations in the $2j + 1$ dimensional space of single nucleon wfs $\psi_{nljm}$. The chain is $U(2j+1) \supset S_p(2j+1) \supset O(3)$, where $U(2j+1)$ and $O(3)$ denote respectively the unitary and orthogonal groups in the corresponding numbers of dimensions, and $S_p(2j + 1)$ denotes the simplectic group in $(2j + 1)$ dimensions, which leaves invariant the antisymmetric state $\left| j^2 J = 0 \right\rangle$. The irreps of $U(2j + 1)$ and $O(3)$ are characterized by the respective quantum numbers T and J, and the irreps of $S_p(2j + 1)$ by the pair of quantum numbers (v, t). The definition of the nuclear seniority scheme for jj-coupling by the above group chain was first given by Flowers[95], and independently by Racah[96].

There are two significant differences between the atomic seniority scheme in LS-coupling and the nuclear seniority scheme in JT-coupling. The first difference is related to the different nature of the interactions: the low atomic states hardly contain [1]S electron pairs due to the strong repulsive interaction of the latter, and the atomic seniority scheme is not based on symmetry (non-symmetry or non-invariance group), and is not a realistic scheme. (See also Hund's rule [TAS 6[7]].) On the other hand, nucleons tend to couple in J = 0 pairs (the single particle model[92]) and the nuclear seniority scheme is based on an approximate symmetry, where the gs and low excited states of spherical nuclei have low seniority[93, 95, 96].

The second difference is related to the additional nuclear quantum number $\beta$. In the atomic case there is no multiplicity corresponding to $\beta$ because the $\frac{1}{2}(n - v)$ added [1]S electron pairs are always in a unique state with L = S = 0, whereas in the nucleus the value of the isospin $t_0$ of the added J = 0 nucleon pairs is variable.

Racah and Talmi[94] introduced the concept of pairing property for identical nucleon interactions which are diagonal in the seniority scheme and their energy matrix for $j^n$ is obtained from that of $j^v$ by adding additively the contribution of the $\frac{1}{2}(n - v)$ added pairs. For such interactions the excitation energy of states of given seniority is n-independent. Racah and Talmi showed that an interaction which is a sum of scalar products of odd-degree irreducible tensors has the pairing property. In particular this is the case for the delta interaction and for (dipole-dipole) Tensor Forces.

### C.  Nuclear masses of lowest seniority states

Edmonds and Flowers studied for $j^n$ configurations the dependence of the energy spectrum on the range of the force for two-nucleon central interactions. They did it by diagonalizing for various values of the range the calculated energy matrices for j values between $j = \frac{3}{2}$ and $j = \frac{7}{2}$ and for (2-3 identical nucleons with) $j = \frac{9}{2}$ [93, 97, 98]. They found that for short range forces the energies increase with T, for a given T they increase with v, and for given T and v they increase when t decreases. Racah[96] reached the same conclusions without matrix diagonalization by elegantly obtaining from group theory an expression for the average energy of a given irrep of $S_p(2j + 1)$ and comparing the averages, which for short range forces are close to the individual levels. Moreover, for the one-dimensional irreps with v = 0 (for even n) and v = 1 (for odd n) he obtained in this way explicit expressions (a mass formula) for the energies of lowest-seniority ground states. This formula is an essential component of the semiempirical lowest-seniority shell model mass equation[99]. He did also a similar calculation for nuclear $1^n$ configurations in LST-coupling[96].

Racah did not apply his equation to the analysis of nuclear masses and spectra. In the nineteen-fifties it was applied by Talmi and his students at the Weizmann Institute of Science, and Talmi described the work in the Rehovoth Conference on Nuclear Structure (1957)[100]. Contrary to Racah's detailed studies of atomic spectra, his work in nuclear physics was on general problems, avoiding detailed analysis of data. Only one of his nuclear physics papers written in Jerusalem addresses a specific spectroscopic problem[90]. Of the M.Sc. and Ph.D. theses in nuclear physics written under his supervision in Jerusalem there are two more M.Sc.[101, 102] and one Ph.D.[103] theses analyzing experimental data, as compared to five M.Sc.[104–108] and three Ph.D.[109-111] theses doing theoretical calculations with no direct application to experiment, and as compared to 46 M.Sc. and 8 Ph.D. theses in atomic spectroscopy analyzing and interpreting experimental spectra (sect. IV.A).

Racah was aware of this point. In his opening lecture as director of the course Nuclear Spectroscopy in Varenna in 1960 he said[112]:

"… since I am more familiar with the mathematics than with the physics of Nuclear Spectroscopy, it was not an easy task for me to organize the programme for this course; however I did my best, which means that I tried to find the best…"
He himself gave the lectures on Mathematical Techniques[64].

When I was his Ph.D. student (1952-1956) I asked Racah about it. He answered that beyond a certain age the human mind is not flexible enough for mastering a new domain and doing creative work in it, whereas in familiar domains one can still continue making significant contributions. We discussed together the theoretical aspects of my work, but for discussing the pertinent experimental data (isomer systematics) I would meet de-Shalit and Talmi in Rehovot.

### D.   New quantum numbers

Ever since the Theory of Complex Spectra Racah continued thinking about the problem of a complete characterization of basis states belonging to irreps of a group chain. He described the motivation for this quest at the Glasgow conference on Nuclear and Mesonic Physics (1954)[113]:

**"A search for new quantum numbers in nuclear configurations**
Giulio Racah
The Hebrew University, Jerusalem

Coefficients of fractional parentage and energy matrices have been calculated in the last few years for many nuclear configurations, and the results were expressed in numerical tables which became bigger and bigger when the configurations became more complex. It is therefore natural to ask whether it is possible to give general formulae for these matrix elements, as functions of the quantum numbers of the states to which they belong. But it is impossible even to try to solve this problem, before we have a *complete* set of quantum numbers for characterizing the different states of a given configuration of equivalent nucleons.

The group-theoretical classification of the states of $1^n$, which was given some years ago (Racah, 1951), still contains three running indices, $\alpha$, $\beta$, $\gamma$ for distinguishing different states having the same conventional quantum numbers; and it is now necessary to substitute these running indices by "quantum numbers", or sets of quantum numbers, which have some physical or mathematical meaning.

It is the purpose of this paper to give some preliminary results obtained in the search for these new quantum numbers."

He reported on partial results for the group chain $U(2\mathbf{l}+1) \supset O(2\mathbf{l}+1) \supset O(3)$ in nuclear $p^n$ and $d^n$ configurations in LST-coupling.

Racah was much occupied[114, 115] with the group chain $U(2j+1) \supset S_p(2j+1) \supset O(3)$ of the nuclear seniority scheme, attempting to overcome the multiplicity problem related to the freedom in choosing the value of $t_0$, the isospin of the added J = 0 nucleon pairs (sect. V.B). He thought of introducing a hermitian operator related to $t_0$ and commuting with the scheme operators, and using its eigenvalues as new quantum numbers. On physical grounds he chose $\mathbf{R} = \sum q_{ik}(\mathbf{t}_i + \mathbf{t}_k)$, but could not find a mathematical connection between $\mathbf{R}$ and $t_0$. Racah's Ph.D. student, Haim Goldberg[110], approached the problem from a group-theoretical point of view and introduced instead of $t_0$ the number $n_\alpha$ of alpha particles in a state with given vtT quantum numbers, where an alpha particle is defined as a four-nucleon antisymmetrized J = T = 0 coupled product of two J = 0, T = 1 two-nucleon states. Heuristically $t_0 = T - t - 2n_\alpha$. The $n_\alpha$ together with vtT make possible a complete definition of scheme, but unlike in quantum mechanics states belonging to different $n_\alpha$ are eigenstates of different non-commuting operators rather than eigenstates of the same operator belonging to different eigenvalues. Using the new scheme for spectroscopic calculations would require the development of new algebraic techniques.

Another group chain studied by Racah together with his Ph.D. students Yehiel Lehrer-Ilamed and Rathindra Nath Sen[111] is $SU(3) \supset O(3)$ used by Elliott for obtaining rotational states in the shell model[116]. Partial results connected with non-orthogonal schemes were reported in the Istanbul Summer School on Group Theoretical Concepts and Methods in Elementary Particle Physics (1962)[117].

## VI. Teaching

Racah was an excellent teacher. He took the subject matter of the courses he taught very seriously and approached it directly, logically and clearly, emphasizing the physical aspects. He was always enthusiastic about what he taught (see also [4, 8, 33, 45]). His lectures at scientific conferences[22, 24, 65, 115, 118], his Varenna[64] and Princeton[43] lectures, and the book Irreducible Tensorial Sets[31] are masterpieces of arrangement of the material and clarity of presentation.

Similarly, Racah always took his students very seriously. He would discuss with them their research problems for hours, and enthusiastically would tell them about the problems he was working on and the results he had obtained. Thus many of the best students were attracted to him, and during the fifteen years when he was the only theoretical physicist in the country (until the arrival of Nathan Rosen at the Technion in Haifa (1954)[119] he established an Israeli school in theoretical physics, which mastered his methods and applied them in various domains. Zvi Lipkin has described the characteristic features of the Israeli school which impressed him upon arrival at the Weizmann Institute in the early nineteen fifties[120]:

"The useful era of quantum mechanics began in the early 50's with the development of the nuclear shell and collective models and of modern solid state physics where quantum mechanics found extensive use as a basic tool without the necessity to solve the deep problems of ultraviolet divergences. During this period I worked at the Weizmann Institute in Israel and was much influenced by the Israeli School of Physics with its pragmatic use of highly sophisticated but down-to-earth theory to describe the real world of experimental data. I should like to thank my friends and colleagues in Rehovoth who helped to teach me in this period that quantum mechanics could be useful and that many apparently complicated and sophisticated formal descriptions were in reality based on very simple physics."

In the introduction to their book on the nuclear shell model [DT (p. 8)], Amos de-Shalit and Igal Talmi, two of Racah's students, have emphasized their indebtedness to him for having introduced them to his methods and for his published as well as unpublished contributions to the subject matter of the book:

"We would only like to refer in particular to the work of G. Racah who can be justly called the founder of modern spectroscopy. The authors are greatly indebted to him for having introduced them to his methods. Much of the material presented in this book is due to him directly as well as indirectly."

## VII. University and Nation-wide Activities

Racah was a strong and active person, with outstanding organizational abilities. He was a leading member in various committees of the university and the senate[121, 122] and held key administrative positions in the university. In his last years he was rector (1961-1965) and acting president (1961-1962)[46]. During his term in office the science faculty which had been scattered in various buildings in Jerusalem since the War of Independence became accommodated together at the Givat Ram campus. A similar transition of the medical school to the Ein Kerem campus was completed. Furthermore, intensive efforts were made to develop all the sciences. Additionally, the University started cultural cooperation with developing countries by teaching the premedical courses to students from Asian and African countries in English [46 (5), 123].

Racah was among the founders of the Israel Academy of Sciences and Humanities (1959) and was the head of the committee that prepared its constitution, and became the first chairman of its Natural Sciences Division (1960-1963). He was among the founders of the Israel Physical Society and its first president (1954-1956). He was a member of The Scientific Council attached to the Prime Minister's Office (later The National Council for Research and Development) from its inception on (1949). He was a member of the Israel Atomic Energy Commission (1952-1955) and a member of Israel delegations to the first three Geneva conferences (1955, 1958, 1964) on the peaceful uses of atomic energy. He was a member of the scientific committee of the Ministry of Defense (1948-1952), and due to his initiative there[124, 125] five of the best physics students (U. Haber-Schaim, G. Yekutieli, I. Pelah, A. de-Shalit and I. Talmi) were sent abroad after the War of Independence to learn nuclear physics and establish nuclear research in Israel in preparation for the developments of the nuclear era. Later (1954)[125] these physicists started nuclear physics studies at the Weizmann Institute, and made it a world center for the nuclear shell model. The group at Rehovot and the department of theoretical physics in Jerusalem kept regular contacts by holding a common weekly seminar at the Weizmann Institute[125] (see also [114]).

In recognition of his scientific achievements Racah became an Honorary Fellow of the Weizmann Institute (1959) and an Honorary Doctor of Science of Manchester University (1961), and was elected an Honorary Foreign Member of the American Academy of Arts and Sciences (1963). He received the Weizmann Prize in the Natural Sciences of Tel Aviv municipality (1954), the Bublick Prize for Science of The Hebrew University (1955), the Israel Prize for Natural Sciences (1958) and the Rothschild Prize (1962). He was also made Commander of the Order of Merit of the Italian Republic (1965). After his death the IAU named a crater on the moon after him (1971)[126]. The Israel Philatelic Service issued a postage stamp of NIS 0.80 with his portrait (1993).

Approaching the end of his term as rector of the university Racah was full of plans for resuming his scientific research which had suffered badly during that period. He intended to attend the Zeeman Centennial Conference on Atomic Spectroscopy in Amsterdam[70] together with his theoretical spectroscopy group (Goldschmidt, Shadmi and Stein). Two days after departing from Israel he died in his family house in Florence from asphyxiation due to a faulty gas heater.

In his lifetime Racah was the only full professor in theoretical physics in Jerusalem. At the end of the period the department included as well an associate professor, a senior lecturer and a lecturer in nuclear physics (G. Rakavy, N. Zeldes and I. Unna), and a lecturer and an instructor in atomic spectroscopy (Y. Shadmi and Z.B. Goldschmidt). In 1970 the departments of experimental and theoretical physics were united as The Racah Institute of Physics.

I owe a debt of gratitude to Issachar Unna for his steady encouragement through the various stages of preparing this article, and to Sam Schweber for his kind help and useful comments during the last stages. Both read the manuscript, and their comments considerably helped improving its accessibility.

I thank Evdokia Alagem, University Archivist, Central Archives of The Hebrew University, Rafael Weiser, director of the Department of Manuscripts and Archives, The Jewish National and University Library, and Elisheva Lahav, coordinator, Hebrew University Photo Archives, for their help in the location of sources.

**Figures**

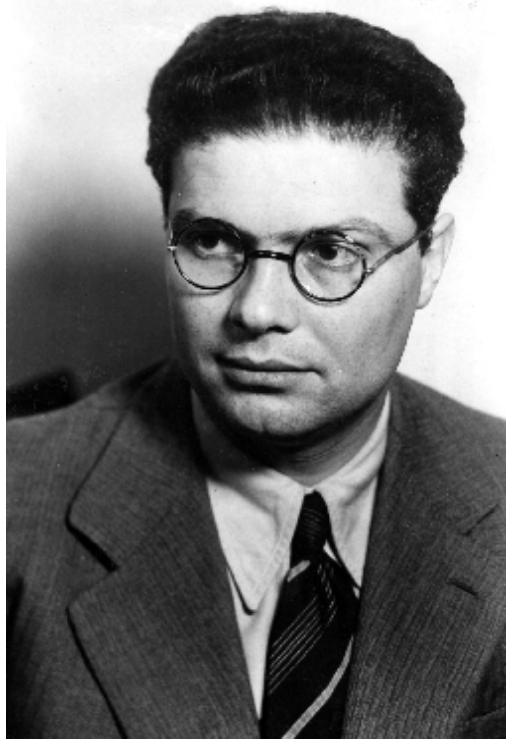

Fig. 1    Professor Racah in his early years in Jerusalem (1942)[127].

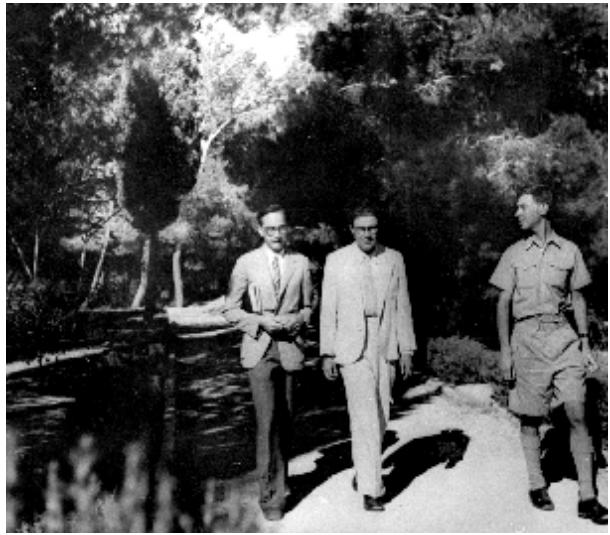

Fig. 2    "Refugee professor" from Italy (center) with assistants, taking a walk on
          campus (1943)[127].

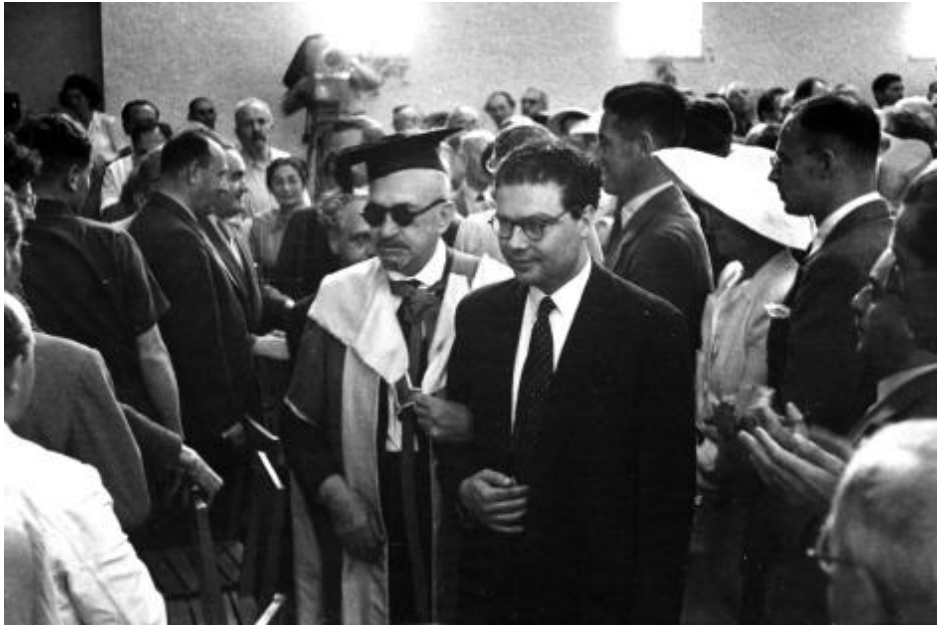

Fig. 3    Dean of Science Racah and Dr. Chaim Weizmann leave the hall after the ceremony of conferring honorary degree of the Hebrew University on Weizmann (27 July 1947)[127].

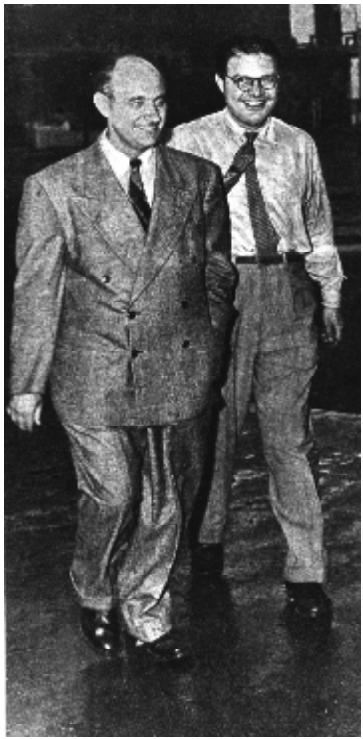

Fig. 4    Racah and Fermi in Rome (1949) [l'Europeo?][46 (2)].

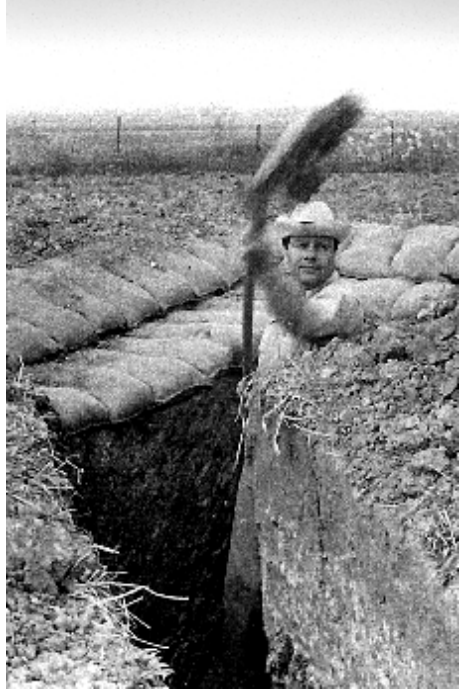

Fig. 5    Racah digging trenches in Kibbutz Deganya Alef during Operation
          Bitzurim, when all classes at the Hebrew University were suspended
          for a week to enable teachers and students to assist in the work of
          building fortifications in the border settlements (29 April – 4 May
          1956) [Courtesy of Dr. Eli Racah].

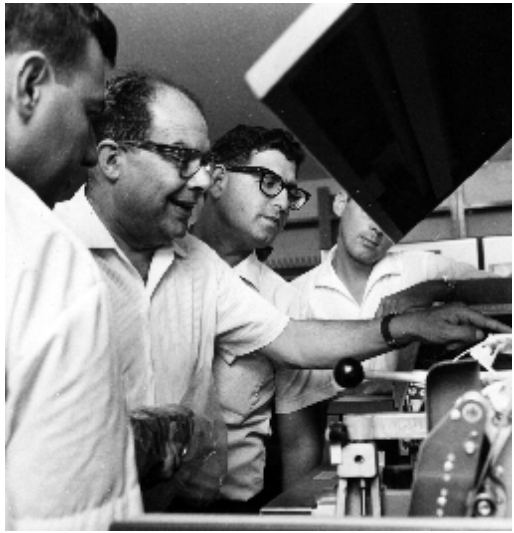

Fig. 7    Last picture of Racah operating the newly acquired IBM 7040
          electronic computer on its arrival at the University (August 1965).
          [Ha'aretz, 6.9.1965] [46 (2)].

La funzione W

Fig. 6   Definition and calculation of the Racah Coefficients (early 1940's)[46 (2)].

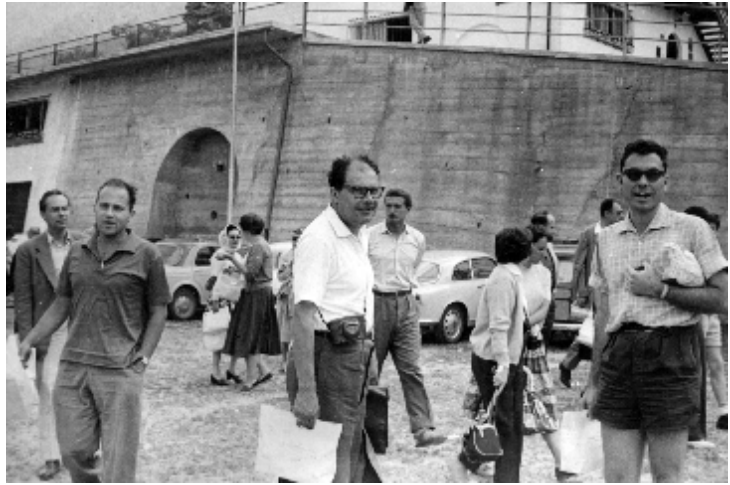

Fig. 8    A weekend excursion near Varenna for the participants of the course Nuclear Spectroscopy of the "Enrico Fermi" international school of physics (June-July 1960) [46 (2)].

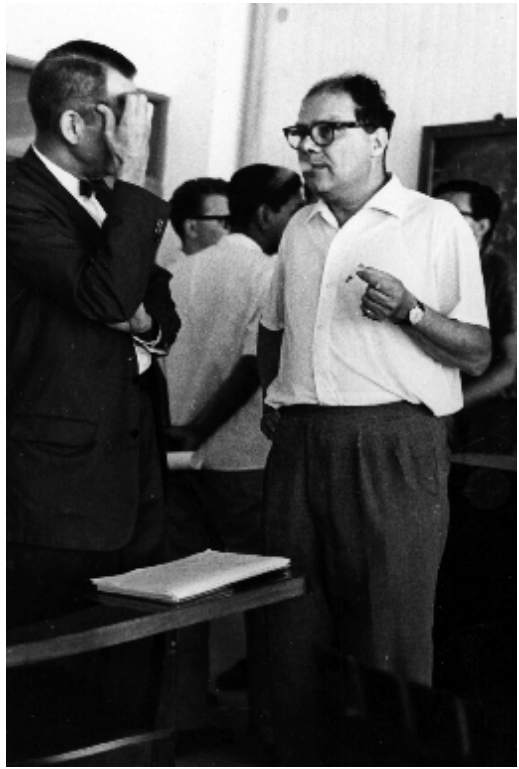

Fig. 9    A coffee break at the Istanbul Summer School on Group Theoretical Concepts and Methods in Elementary Particle Physics (July-August 1962) [46 (2)].

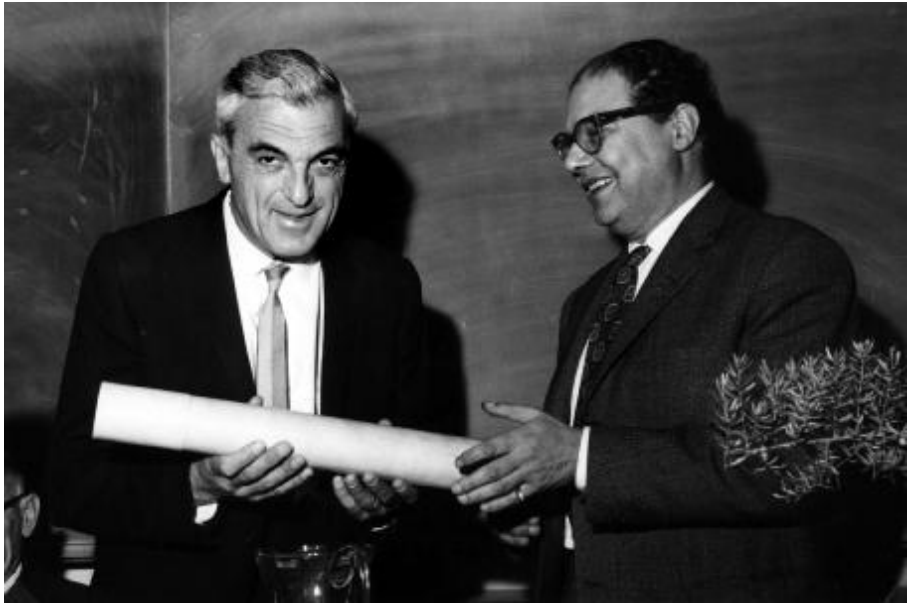

Fig. 10    Rector and Acting President Racah conferring honorary doctorate of
the Hebrew University on Prof. Felix Bloch (February 1962)[127].

.